\documentclass[preprint]{aastex}

\shorttitle{Ensemble Blazar Variability in the P-Q Survey}
\shortauthors{Bauer et al.}

\begin{document}

\begin{titlepage}

\title{Blazar Optical Variability in the Palomar-QUEST Survey}
\author{Anne Bauer\altaffilmark{1,2}, Charles Baltay\altaffilmark{1}, Paolo Coppi\altaffilmark{1}, Nancy Ellman\altaffilmark{1}, Jonathan Jerke\altaffilmark{1}, David Rabinowitz\altaffilmark{1}, Richard Scalzo\altaffilmark{1}}
\email{anne.bauer@aya.yale.edu}
\altaffiltext{1}{Yale University, Department of Physics, P.O. Box 208120, New Haven, CT 06520-8120, USA}
\altaffiltext{2}{Universit\"{a}ts-Sternwarte M\"{u}nchen, Scheinerstr. 1, D-81679 M\"{u}nchen, Germany}
\begin{abstract}

We study the ensemble optical variability of 
276 FSRQs and 86 BL Lacs in the Palomar-QUEST Survey with the goal of searching for 
common fluctuation properties, examining the range of behavior across 
the sample, and characterizing the appearance of blazars in such a survey 
so that future work can more easily identify such objects.  The survey, 
which covers 15,000 square degrees multiple times over 3.5 years, 
allows for the first ensemble blazar study of this scale.  
Variability amplitude distributions are shown 
for the FSRQ and BL Lac samples for numerous time lags, and also studied 
through structure function analyses.  
Individual blazars show a wide range of variability amplitudes, 
timescales, and duty cycles.  Of the best sampled objects, 35\% are 
seen to vary by more than 0.4 magnitudes;  for these, the fraction 
of measurements contributing to the high amplitude variability ranges 
constantly from about 5\% to 80\%.  
Blazar variability has some similarities 
to that of type I quasars but includes larger amplitude 
fluctuations on all timescales.  
FSRQ variability amplitudes are particularly similar to those of 
QSOs on timescales of several months, suggesting significant 
contributions from the accretion disk to the variable flux at these 
timescales.  Optical variability amplitudes are correlated with the 
maximum apparent velocities of the radio 
jet for the subset of FSRQs with MOJAVE VLBA measurements, implying 
that the optically variable flux's strength is 
typically related to that of the radio emission.  
We also study CRATES radio-selected 
FSRQ candidates, which show similar variability characteristics to 
known FSRQs;  this suggests a high purity for the CRATES sample.

\end{abstract}


\maketitle

\end{titlepage}

\section{Introduction}

Blazars are among the most variable objects known.  Fluctuations 
occur on all observed timescales, and shifts of several magnitudes 
in optical brightness have been seen over the span of just months 
(see, e.g., \cite{metcalf06}).  The mechanisms behind these dramatic 
fluctuations 
are not well understood.  A number of individual blazars have been 
carefully observed, revealing a large range of variability behavior.  
(e.g. \cite{chatterjee08}; \cite{raiteri08}).  
Characteristic fluctuation timescales measured for specific blazars 
range from hours to years (e.g. \cite{heidt96}; \cite{ciprini03}; 
\cite{kartaltepe07}).  

Traditionally, blazar variability has been studied through intensive 
monitoring of individual objects.  With new wide-field surveys 
it is possible to take a complementary approach by studying a few 
measurements of many blazars.  If the blazar class is completely 
homogeneous, the ensemble properties will be representative of each 
object.  If the class is heterogeneous, then the ensemble properties 
will correspond exactly to none of the objects, but will nevertheless 
be informative in statistically describing the sample.  The ensemble 
variability of type I quasars has been studied in this manner 
(e.g. \cite{vandenberk04}; \cite{quasar_paper}) and compared with more 
detailed work on individual objects (e.g. \cite{collier01}) in order 
to understand better the causes of quasar variability and the range of 
physical processes at work in these systems.  Observations of individual 
AGN suggest that blazar variability exhibits much more diversity of 
amplitude and timescale than quasar variability.  
No large ensemble study of blazar
variability has yet been published to study this effect, however, 
due to the lack of repeated measurements over large enough areas of sky.  
Wide-field surveys such as Palomar-QUEST are able to observe large 
numbers of blazars and begin to study the collective variability 
properties of the class, and to examine the extent of heterogeneity within 
the sample.  

The characterization of blazar variability as seen by several measurements 
over a wide-field survey is useful for another purpose:  as a tool 
for using variability to identify blazars in such surveys.  Knowledge of 
blazar physics is limited in part because relatively few blazars are 
known;  therefore the discovery of more blazars is important for a better 
understanding of the objects' behavior.  
Furthermore, nearly all blazars currently studied have been 
selected by their flux ratios in the radio and/or X-ray bands, possibly 
contributing to significant selection effects in the ensemble.  As well 
as typically exhibiting strong radio and X-ray emission, blazars are 
also some of the most optically variable objects 
in the sky.  By studying the appearance of blazars as a class in 
sparsely sampled variability surveys we can perhaps facilitate the 
discovery of many more in current surveys like Palomar-QUEST, as well 
as imminent surveys such as Pan-STARRS.

Blazars are currently understood to be active galactic nuclei (AGN) with 
jets, where the jet axis is aligned along the observer's line of sight.  
Their flux is therefore dominated by beamed jet emission 
across the electromagnetic spectrum.  
The broadband spectral energy distributions (SEDs) of blazars 
exhibit two broad peaks, one at lower energies (ranging in 
frequency from radio to X-rays) and one at higher energies (from X-rays 
to TeV gamma rays).  The low-energy peak 
is commonly attributed to synchrotron emission from high-energy 
particles accelerated in regions of shocked gas in the jet.  These 
high-energy particles can also reprocess synchrotron radiation from the 
low-energy peak, thermal photons from the accretion disk, or other 
external photons via Compton upscattering to form the high-energy peak.  
Hadron synchrotron radiation or decay may also contribute to the 
high-energy peak (see, e.g., \cite{mucke01}).

Blazars are often 
divided into subcategories of BL Lac objects and flat spectrum radio 
quasars (FSRQs), based in part on the absence or presence, respectively, 
of broad optical emission lines superimposed on the characteristic jet 
continuum emission.  These broad lines are accretion disk emission 
reprocessed by energetic gas that is close to the system's central black 
hole.  These emission lines are visible in FSRQs but are most 
prominent in type I quasars, which are AGN aligned somewhat off-axis such 
that the observer views flux predominantly from the accretion disk.  
Reviews of the typical constituent parts of an AGN and the orientation-based 
theory of AGN unification linking quasars and blazars can be found in, for 
example, \cite{urry} and \cite{peterson}.

FSRQs and BL Lacs are believed to differ not just in their emission 
lines, but also in the structure and power of their jets.  
Radio galaxies, or AGN with misaligned jets, are divided into two 
basic categories:  Fanaroff-Riley types I and II (FR I and FR II)
(\cite{fr74}).  FR I jets generally have lower energies
and are spatially limited, with the power concentrated close to the
center of the galaxy.  FR II jets are more energetic and extend much
further out, with the most luminous areas in lobes far from the
center of the galaxy.  FSRQs and BL Lacs are believed to belong to 
the same population as FR IIs and FR Is, respectively.  FSRQs are more 
luminous than BL Lacs, as FR IIs are compared to FR Is.  Furthermore, 
the relative number counts of FSRQs with respect to FR IIs, 
and BL Lacs with respect to FR Is, are consistent with the blazars 
being those radio galaxies that are by chance aligned to within 
$\sim 15^{\circ}$ of the line of sight (\cite{urry}).  
This blazar/radio galaxy unification model implies that FSRQs have 
higher energy jets than BL Lacs.  

Because of the morphological, energetic, and spectral differences 
between FSRQs and BL Lacs, we study the two types separately.  
Distinctions between the optical variability of the groups can 
improve our understanding of the fundamental differences between 
these subclasses.

The dramatic variability of blazar flux has several possible causes.  
For example, the intermittent nature of shocks in the jet produces 
highly variable emission.  
Much work has been done to model shock dissipation mechanisms in order 
to understand blazars' variability and energy distributions.  However, 
current models are not well constrained;  even a single model framework 
can yield many possible variability characteristics (see, e.g., \cite{li00}).  
It is also uncertain 
how the sources of different wavelength emission are physically related 
to each other, for example if the radio, optical, and/or X-ray flux 
are generated in the same jet location.  Multiwavelength monitoring of 
several blazars has revealed some cases in which radio, optical, 
and X-ray flares appear to have a common source;  other cases 
show one band's emission to be unaffected by another's fluctuations 
(see, e.g., \cite{darcangelo07};  \cite{chatterjee08};  \cite{marscher08};  
\cite{bonning09}).

Variability may also be due to geometric effects.  Shocks and instabilities 
can kink the jet, changing the beaming direction with respect to our 
line of sight and therefore affecting how much jet flux reaches us.  
Or, beaming direction changes can be due to the 
precession of the black hole if it is in a binary system.  Precession is not 
thought to be a dominant source of variability, as most blazars studied 
have clearly aperiodic flux patterns.  Some blazars, however, do show 
somewhat regular variability.  \cite{ciaramella04} 
found periodic features in the radio emission of 5 out of 77 blazars 
studied, with periods ranging from about 3 to 9 years.  

The Palomar-QUEST 
Survey scans 15,000 square degrees of sky multiple times, 
thereby repeatedly observing a large number of blazars in optical 
wavelengths.  Those with consistent classification, precise 
redshift measurements, and multiple good-quality Palomar-QUEST 
measurements number 362 in total: 86 BL Lacs and 276 FSRQs.  Each 
blazar has, on average, roughly four observations.  By compiling 
and comparing sparse variability measurements of an unprecedented 
number of blazars we can study the typical variability behavior of 
the blazar class and the range of common fluctuations.

In \S \ref{survey_section} we briefly describe the Palomar-QUEST 
Survey.  \S \ref{sample_section} introduces the samples 
of BL Lacs and FSRQs used in this analysis. \S \ref{analysis_section} 
discusses the techniques which we use to analyze timescales 
of variability.  Results are 
presented in \S \ref{results_section} 
and discussed in \S \ref{discussion_section}, and we conclude in 
\S \ref{conclusions_section}. 

\section{The Palomar-QUEST Survey \label{survey_section}}

The Palomar-QUEST Survey is a large area optical survey which uses 
the 48'' Samuel Oschin Schmidt Telescope at Palomar Observatory.  Over 
the span of 3.5 years we have observed 15,000 square degrees of sky 
multiple times with seven optical filters: Johnson UBRI and SDSS 
r'i'z'.  
The sky area has been covered typically 4-5 times in each 
filter, with the time between passes ranging from hours to years.
The data were taken using the QUEST2 Large Area Camera, which was built 
for the survey.  The camera is made up of 112 CCDs, covering in total 
$4.6^{\circ} \times 3.6^{\circ}$, which are arranged into 4 rows such 
that a separate filter can be placed over each row.  The camera 
was used in driftscan 
mode to take the data.  In this mode, an 8 hour night of observing 
yields $\sim500$ square degrees of data in 4 filters with an exposure 
time of roughly 140 seconds.  The camera is described in detail in 
\cite{camera_paper}.  

The Palomar-QUEST data are processed using custom-written software.  
Objects are detected by fitting flux peaks with an empirical point 
spread function (PSF) model, which allows for accurate treatment of 
close neighbors.  Photometry is performed using PSF and also aperture 
fitting, with the PSF measurements used as the primary results.  
Astrometry is calculated with respect to the USNO A-2.0 catalog 
(\cite{usno}), and is accurate and precise to 0.1''.  This  processing 
software is described in detail in \cite{software_paper}.  

The calibration 
of the Palomar-QUEST data for variability work involves five basic steps.  
First, zeropoints are applied to correct for sensitivity variations 
across each individual chip as well as for nonlinearities in each 
chip, and also for different response levels from chip to chip.  
These zeropoints were calculated once, and correct for the constant 
response characteristics of the CCDs.  Next, for each 1/16 square 
degree of sky under consideration, a photometrically stable Palomar-QUEST 
scan is chosen to be used as the calibration standard.  
An RA-dependent correction is 
applied to all scans overlapping the standard to account for changing 
weather conditions.  A declination-dependent calibration corrects for 
spatially-dependent factors such as scattered light in the camera.  
Finally, strict quality cuts are made to eliminate poor measurements 
and substandard data regions.  Because it is important for variability work 
to have as many comparable measurements of an object as possible, we 
calibrate together data from the Johnson R and SDSS r' filters to form 
an ``Rr'' bandpass.  Since the R and r' wavelength ranges are similar, 
the color terms introduced by this combination are small.  The calibration 
yields a systematic error of 0.7\% for typical pointlike objects in the 
``Rr'' band, which are the data used in this work.  For a 
thorough explanation of the variability calibration see \cite{quasar_paper}, 
referred to henceforth as Paper I.  

\section{The Blazar Sample \label{sample_section}}

We have studied a set of blazars accumulated from various sources: 
\cite{1jansky}, \cite{hewitt93}, \cite{collinge05}, \cite{donato05}, 
\cite{gamma1}, \cite{vcv}, \cite{massaro07}, 
\cite{roxa}, and \cite{cgrabs}.  
These lists have many objects in common.  In total, 
in the Palomar-QUEST Survey area, including only objects whose 
positions are known to roughly 1 arcsecond, whose redshifts have 
been measured, and which have multiple well-calibrated Palomar-QUEST 
observations, we have collected 94 unique BL Lacs and 278 unique 
FSRQs.  There are also four objects that were classified by some studies
as BL Lacs and by others as FSRQs.  This highlights the fact that a
blazar's broad emission lines often vary in prominence over time,
sometimes to the point of making the classification unreliable.
We did not include the four ambiguous objects in our study.  There
may be more such blazars that remain in the sample, thereby confusing
the results.  Since only about 1\% of the sample is doubly classified,
however, we expect the effect to be small.

\begin{figure}
\begin{center}
\plotone{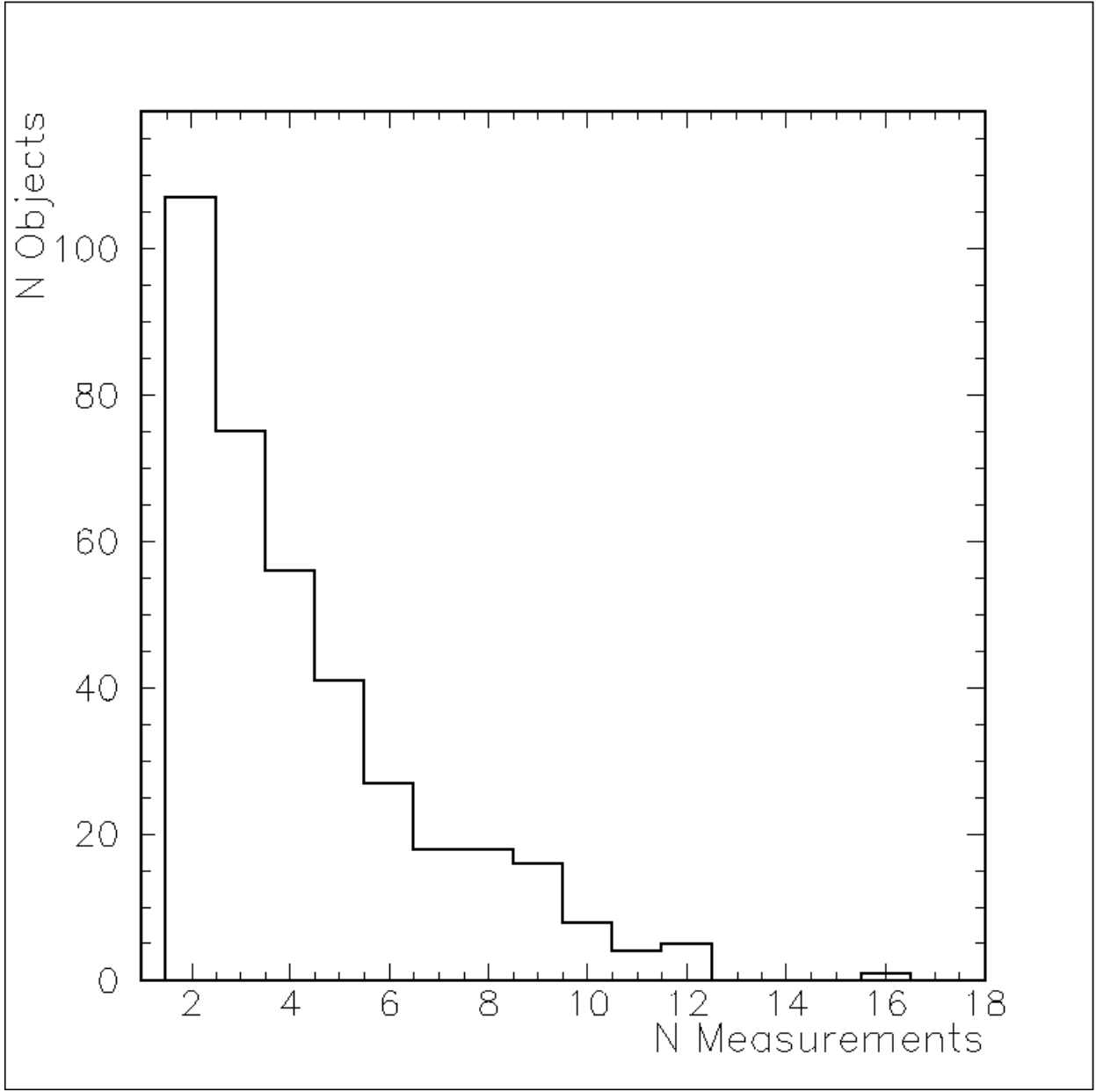}
\caption{Number of blazars with N good Rr measurements.}
\label{nhistogram}
\end{center}
\end{figure}

\begin{figure}
\begin{center}
\plotone{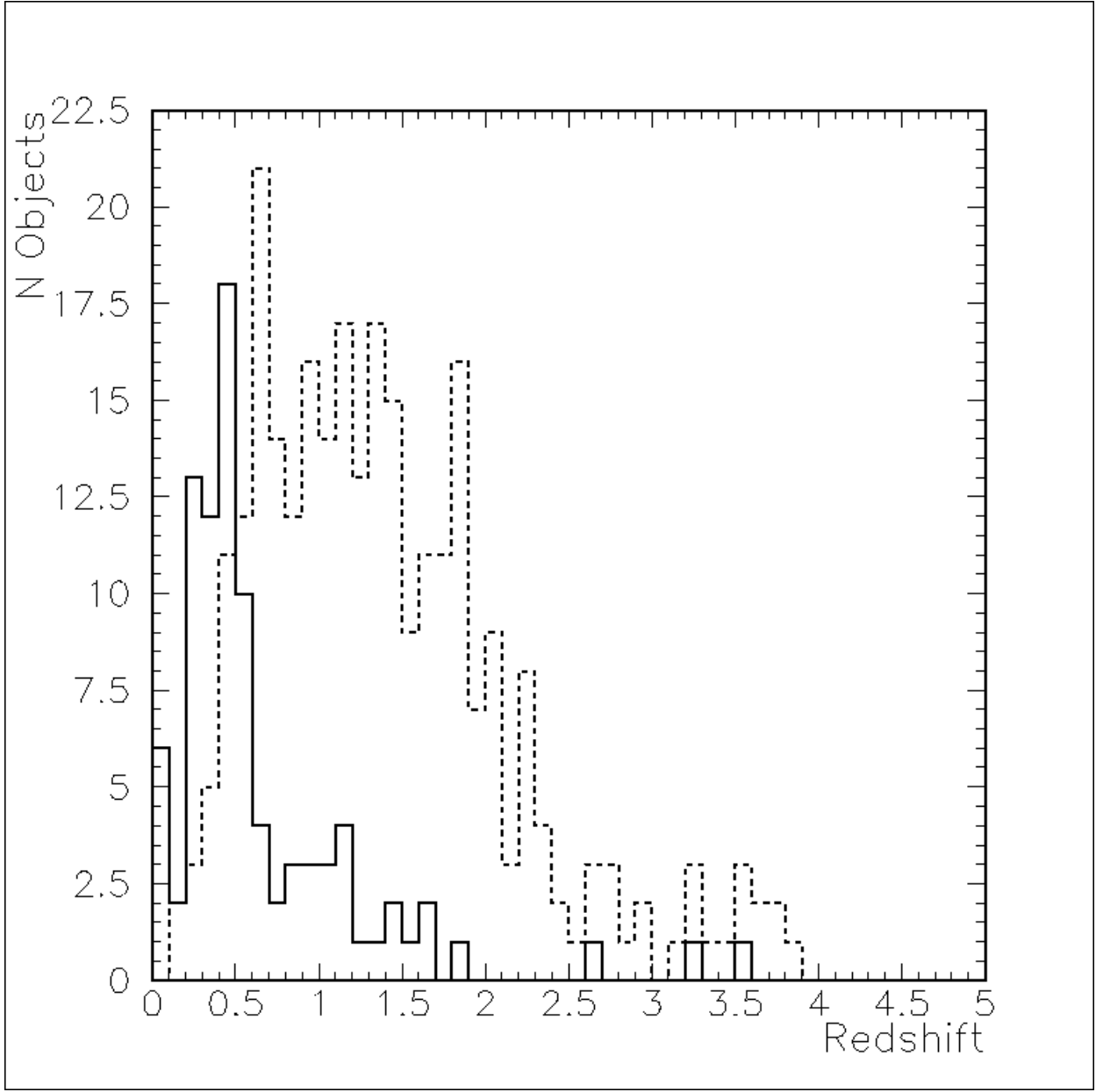}
\caption{Redshift distribution of the blazars.  Solid line: BL Lacs.  Dashed line: FSRQs.}
\label{blazar_redshift}
\end{center}
\end{figure}

A histogram of the number of good quality QUEST Rr observations of each 
blazar is shown in figure \ref{nhistogram}.  
The redshift distribution of the blazar sample is shown in figure 
\ref{blazar_redshift}.  Note the difference between the BL Lac and 
FSRQ redshift distributions, due to the fact that BL Lacs are intrinsically 
fainter than FSRQs, and therefore can be detected by QUEST only at lower 
redshifts (see, e.g., \cite{padovani07}).  The blazar distribution 
is suppressed at redshifts lower than 0.2.  The catalogs 
from which we took the objects do not have this limit;  there are many 
blazars, particularly BL Lacs, with lower redshifts.  However, the 
variability calibration of the Palomar-QUEST data rejects measurements 
of objects that appear spatially extended, as our flux measurements of 
extended objects 
are not reliable.  The details of and motivation for this cut are discussed 
in Paper I.  Most of the blazars with redshifts below 0.2 
appear extended in our data and therefore do not yield good quality 
measurements.  
We therefore remove all blazars from our 
list that have redshifts less than 0.2.  This cut eliminates 8 BL Lacs and 
2 FSRQs, leaving us with a final sample of 86 BL Lacs and 276 FSRQs.

\section{Analysis Techniques \label{analysis_section}}

\subsection{SF($\tau$) \label{sf_section}}

The structure function is a statistical quantity often used to study 
the variability of AGN (see, e.g. Paper I, 
\cite{vandenberk04}).  It can be defined as

\begin{equation}
SF(\tau) = \sqrt{<[m(t) - m(t-\tau)]^{2}> - <\sigma^{2}>}
\end{equation}

where $m(t) - m(t-\tau)$ is the difference in measured magnitudes of 
an object at two epochs separated by rest frame 
time $\tau$.  $\sigma$ is the measurement error on the magnitude 
difference term.  $<X>$ denotes the mean value of $X$ over the set of 
objects.  We perform three iterations of $3\sigma$ clipping prior 
to averaging the data.  
The structure function is a measure of the intrinsic 
variability of an object, and as it is a function of rest frame 
time lag $\tau$ it 
is useful in determining characteristic timescales of variability.  

The shape of the structure function vs. $\tau$ depends 
on the sample's variability in certain characteristic ways.
For short timescales, if the intrinsic variability is less 
than the measurement noise, then the structure function will 
plateau.  For timescales longer 
than the maximum timescale characteristic of the variability, the 
plot will also plateau.  
Between these possible plateaus the structure function rises with a 
shape determined by the details of the system.  For example, if the 
variability is characterized by a power law frequency distribution then 
the structure function will also yield a power law shape, with an index 
directly related to the frequency dependence.  Other variability 
patterns can produce more complicated structure functions.  
In general, changes in the shape of the structure function indicate 
timescales important to the variability mechanism.  

\subsection{V($\tau$)}

Because it averages over variability amplitudes, the structure 
function does not describe the range of variability behavior observed in 
a sample.  If two samples have a similar structure 
function it is not clear if the distributions of 
variability amplitudes are truly similar, or if they only have similar 
mean values.  To show the range of blazar behavior explicitly, we 
calculate, for each pair of measurements for each object, 
\begin{equation}
V(\tau) = \sqrt{[m(t) - m(t-\tau)]^{2} - \sigma^{2}}
\end{equation}
where $m(t)$ and $m(t-\tau)$ are two magnitude measurements of the same 
object separated by rest frame time lag $\tau$, and $\sigma$ is the 
uncertainty on the magnitude term.  This 
quantity is similar to the structure function except it is not an 
average over the sample; a blazar with 
$N$ magnitude points will yield $N \times (N-1)/2$ 
separate values of $V$.  For measurement pairs where no variability 
is seen and $[m(t) - m(t-\tau)]^{2} < \sigma^{2}$, $V$ is set to zero.

\section{Results \label{results_section}}

\subsection{Distributions of V($\tau$)}

\begin{figure}
\begin{center}
\plotone{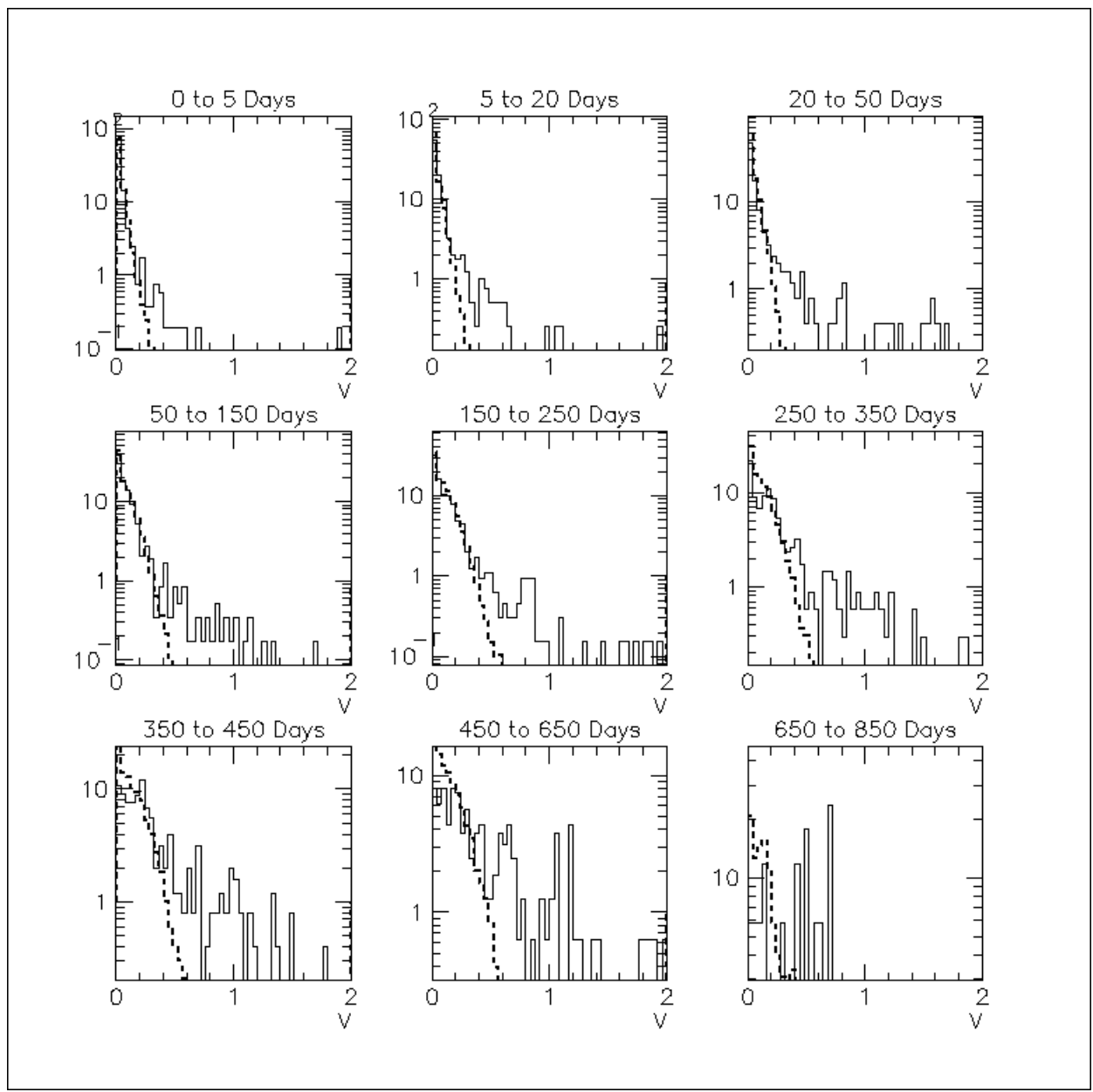}
\caption{Variability $V$ histograms of FSRQs (solid line) and quasars (dashed line) for 9 different time lag bins.}
\label{fsrqqso}
\end{center}
\end{figure}

\begin{figure}
\begin{center}
\plotone{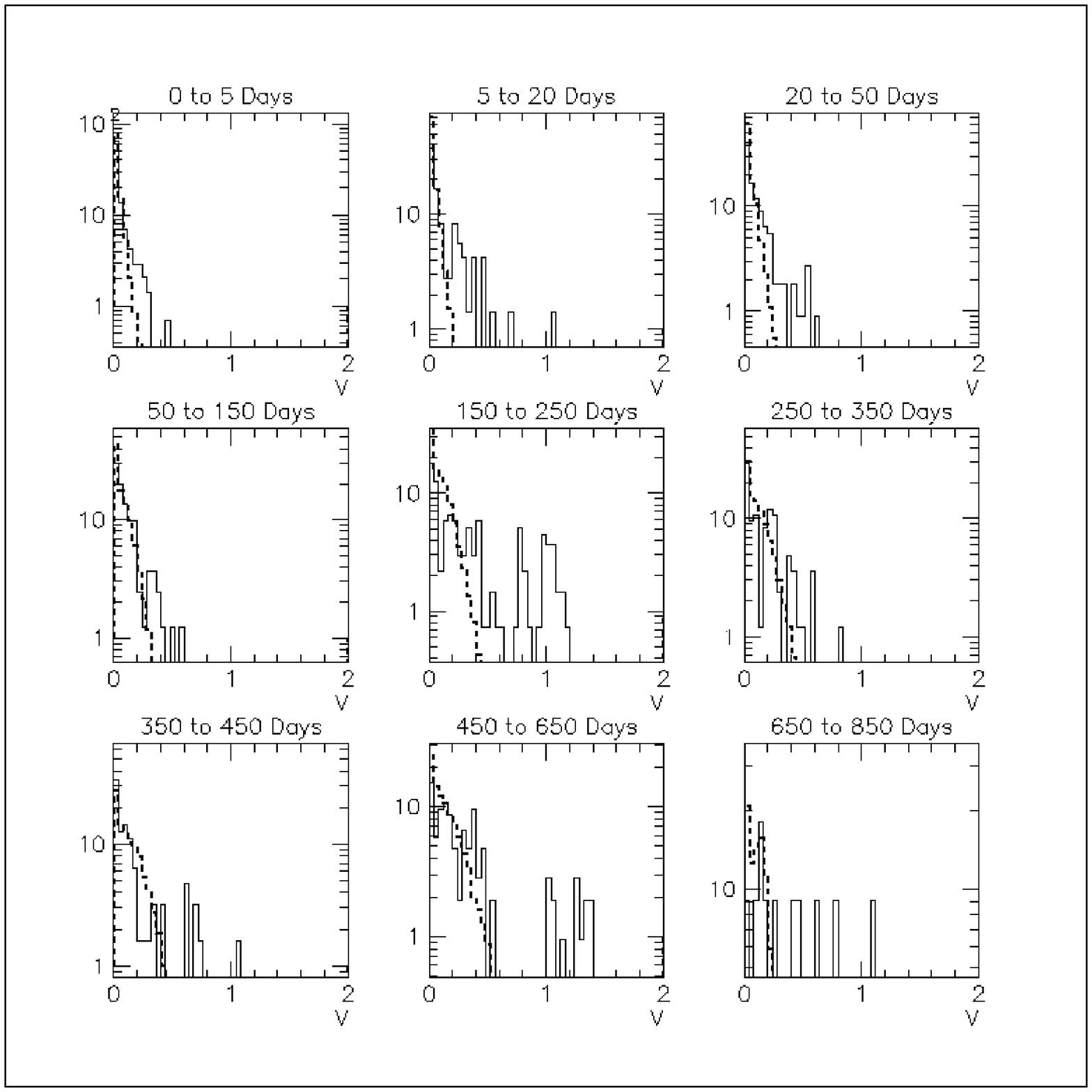}
\caption{Variability $V$ histograms of BL Lacs (solid line) and quasars (dashed line) for 9 different time lag bins.}
\label{bllacqso}
\end{center}
\end{figure}

Figures \ref{fsrqqso} and 
\ref{bllacqso} show histograms of $V$ for FSRQs and BL Lacs, 
respectively, each split into nine $\tau$ bins.  
Also shown in each plot as a dashed line are the corresponding data 
for the type I quasar 
sample studied in Paper I.  The blazar and quasar histograms are each 
normalized to a total of 100 objects so that the two are easily 
comparable by eye.

These plots show qualitative differences between the samples.  
Both BL Lacs and FSRQs show excess high amplitude variability compared 
to the quasars on all timescales.  
Both BL Lacs and FSRQs often show a low amplitude variability peak 
similar to that of quasars, particularly at shorter timescales.  
This quasar-like peak is more pronounced in FSRQs than in BL Lacs.
The FSRQs show more variability at the highest 
amplitudes, with $V \gtrsim 1$ magnitude, than the BL Lacs.  

\subsubsection{Effects of Sampling on $V(\tau)$ \label{v_sampling_section}}

Repeated QUEST measurements of each blazar in the sample are not 
made at regular intervals.  For example, sometimes measurements were 
made of the same sky area three months apart.  However, it is 
impossible to observe the same sky area from the ground at a six 
month interval.  Because each of the nine $\tau$ bins over which we 
measure $V$ are not populated by measurements with a uniform $\tau$ 
distribution, the results in figures \ref{fsrqqso} and \ref{bllacqso} 
may be skewed by the details of our sampling cadence.  To study this 
effect, we have examined how the data's windowing function alters 
$V$ histograms measured from one blazar with unusually good historical 
sampling: 3C 273.  
Ideally, we would like to study the effects of sampling on a 
``typical'' blazar lightcurve.  However, sufficient data have not 
been collected for enough individual blazars over the timescale of 
the QUEST survey to empirically determine ``typical'' behavior.  
We do not wish to interpolate between sparse measurements of a 
large number of blazars to calculate an averaged lightcurve, as 
an interpolation between distant points would not reproduce the 
short term variability we know to exist in blazars.  
Furthermore, theoretical models of synchrotron emission lack the 
short-term variability commonly observed in blazar lightcurves 
(see, e.g., \cite{lindfors06}), and are therefore not sufficient 
as models for 
studying how sampling affects the measurement of such variability.  
We therefore take 3C 273, a blazar with exceptional coverage 
over several decades, as an example object to probe the effects 
of windowing in the QUEST data.  
Because the data cadence's effects on a measurement depend on the 
details of objects' lightcurves, one blazar is not sufficient to 
completely describe the effects of windowing on our sample.  
However, if the results are strongly and systematically skewed 
by our uneven data cadence, we expect every blazar lightcurve to 
be influenced by the effect.  3C 273 should therefore be a 
useful diagnostic to detect significant windowing effects in the 
results.  For this work we use V-band data available at 
http://isdcul3.unige.ch/3c273.

\begin{figure}
\begin{center}
\plotone{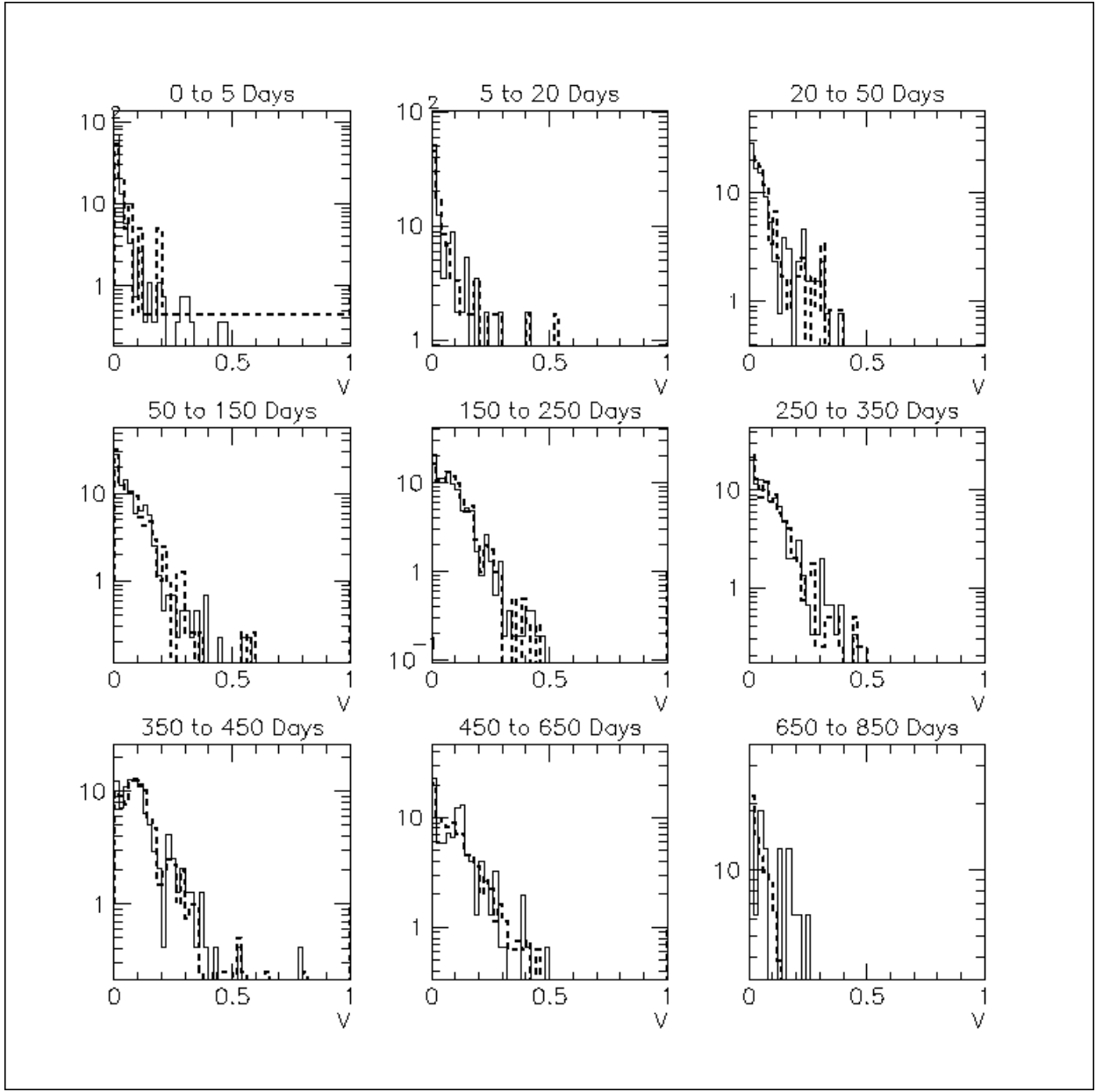}
\caption{$V$ histograms for 3C 273 using data evenly distributed in $\tau$ (dashed line) and data distributed with the QUEST FSRQ sampling cadence (solid line).}
\label{fsrq_windowed}
\end{center}
\end{figure}

\begin{figure}
\begin{center}
\plotone{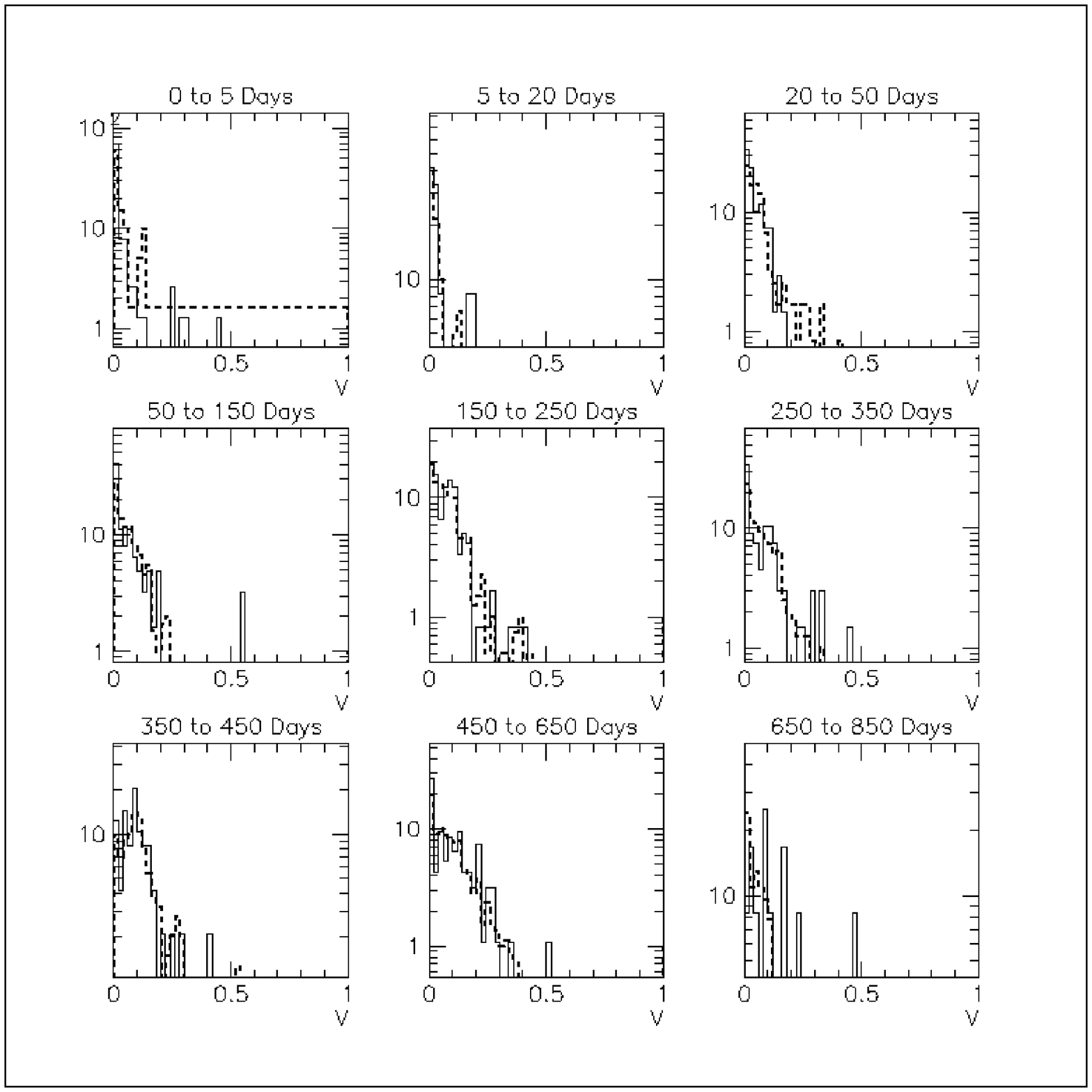}
\caption{$V$ histograms for 3C 273 using data evenly distributed in $\tau$ (dashed line) and data distributed with the QUEST BL Lac sampling cadence (solid line).}
\label{bllac_windowed}
\end{center}
\end{figure}

To construct $V$ histograms from 3C 273 data that are 
not affected by windowing, we use a subset of the 
available measurements which are distributed nearly evenly in 
$\tau$.  First, $V$ is calculated for each pair of measurements 
in the lightcurve.  This yields a large number of $V$s with a 
non-uniform distribution of $\tau$ ranging from hours to nearly 
40 years.  We then bin the $\tau$ values 
in units of one day (bins of less than 1 day would not all be 
populated since the measurements do not have spacings of $\sim$0.5 
day).  From each bin we randomly choose a $V$ measurement with which 
to populate the evenly-sampled V histograms.  
The data are 
split into nine $\tau$ histograms as in figures \ref{fsrqqso} 
and \ref{bllacqso} in order to mimic the QUEST results.  For 
all but the first $\tau$ histogram, the one-day binning in $\tau$ 
introduces a small $\tau$ error compared to the length of the 
measurements' time lags.  For the 0-5 day $\tau$ histogram, this 
procedure will not probe the effects of the data cadence as 
finely, as each bin width comprises 20\% of the $\tau$ range.  $V$ 
histograms constructed using the nearly evenly-distributed 
3C 273 data are shown as the thick dashed lines in figures 
\ref{fsrq_windowed} and \ref{bllac_windowed}.  

To assess the effects of the QUEST data cadence, we 
recalculate the $V$ histograms for a subset of the 3C 273 data 
which adheres to the QUEST sampling rate.  The 
$\tau$ distribution for the QUEST blazar measurements is 
calculated and binned into one-day bins.  
For each one-day bin, for each QUEST measurement pair we 
select at random a $V$ from the same $\tau$ bin of 3C 273 data.  
$V$ histograms constructed using this QUEST-sampled 3C 273 dataset 
are shown as the solid lines in figures \ref{fsrq_windowed} and 
\ref{bllac_windowed}:  the FSRQ sampling is used for figure 
\ref{fsrq_windowed}, and the BL Lac for figure \ref{bllac_windowed}.  
In both figures, both datasets are normalized to a total of 100 
objects in order to be easily comparable by eye.

Figures \ref{fsrq_windowed} and \ref{bllac_windowed} show that the 
evenly-distributed and QUEST-distributed 3C 273 data yield 
similar $V$ histograms.  The exact histogram shapes depend on the 
random $V$ values chosen from each 3C 273 bin.  To quantitatively compare 
the distributions we create 100 such evenly-sampled and 
QUEST-sampled sets and compare them, separately for each $\tau$ 
interval, using a Kolmogorov-Smirnov (K-S) test.  For this comparison we 
ignore measurement pairs with $[m(t) - m(t-\tau)]^{2} < \sigma^{2}$ rather 
than setting $V = 0$ in those cases, as the K-S test is sensitive to the 
shape of the distribution and will be strongly affected by an 
artificially sharp peak at $V = 0$.  
The 100 K-S 
results for each $\tau$ histogram 
are averaged to find a best estimate of the probability that 
the QUEST-sampled histograms come from the same underlying distribution 
as the evenly-sampled histograms.  The results of the K-S tests 
are shown in table \ref{window_ks_table}.  It is not possible to 
distinguish the two datasets as coming from different distributions.  
We therefore conclude that the details of the QUEST sampling cadence 
do not significantly affect the shapes of the $V$ histogram results 
for our sample blazar.

\begin{table}
\begin{center}
\begin{tabular}{|l|l|l|}
\hline
$\tau$ (Days) & FSRQ P & BL Lac P \\\hline
0-5 & 0.33 & 0.48 \\ 
5-20 & 0.43 & 0.38 \\ 
20-50 & 0.48 & 0.53 \\ 
50-150 & 0.49 & 0.58 \\ 
150-250 & 0.59 & 0.52 \\ 
250-350 & 0.46 & 0.49 \\ 
350-450 & 0.59 & 0.56 \\ 
450-650 & 0.27 & 0.39 \\ 
650-850 & 0.52 & 0.48 \\\hline 
\end{tabular} 
\end{center}
\caption{K-S probabilities P that the $V$ histograms from evenly-distributed and QUEST-distributed 3C 273 data come from the same underlying distribution.  Calculated separately for FSRQ and BL Lac sampling cadences.}
\label{window_ks_table}
\end{table}

\subsection{SF($\tau$) vs. $\tau$ \label{sf_vs_time_section}}

\begin{figure}
\begin{center}
\plottwo{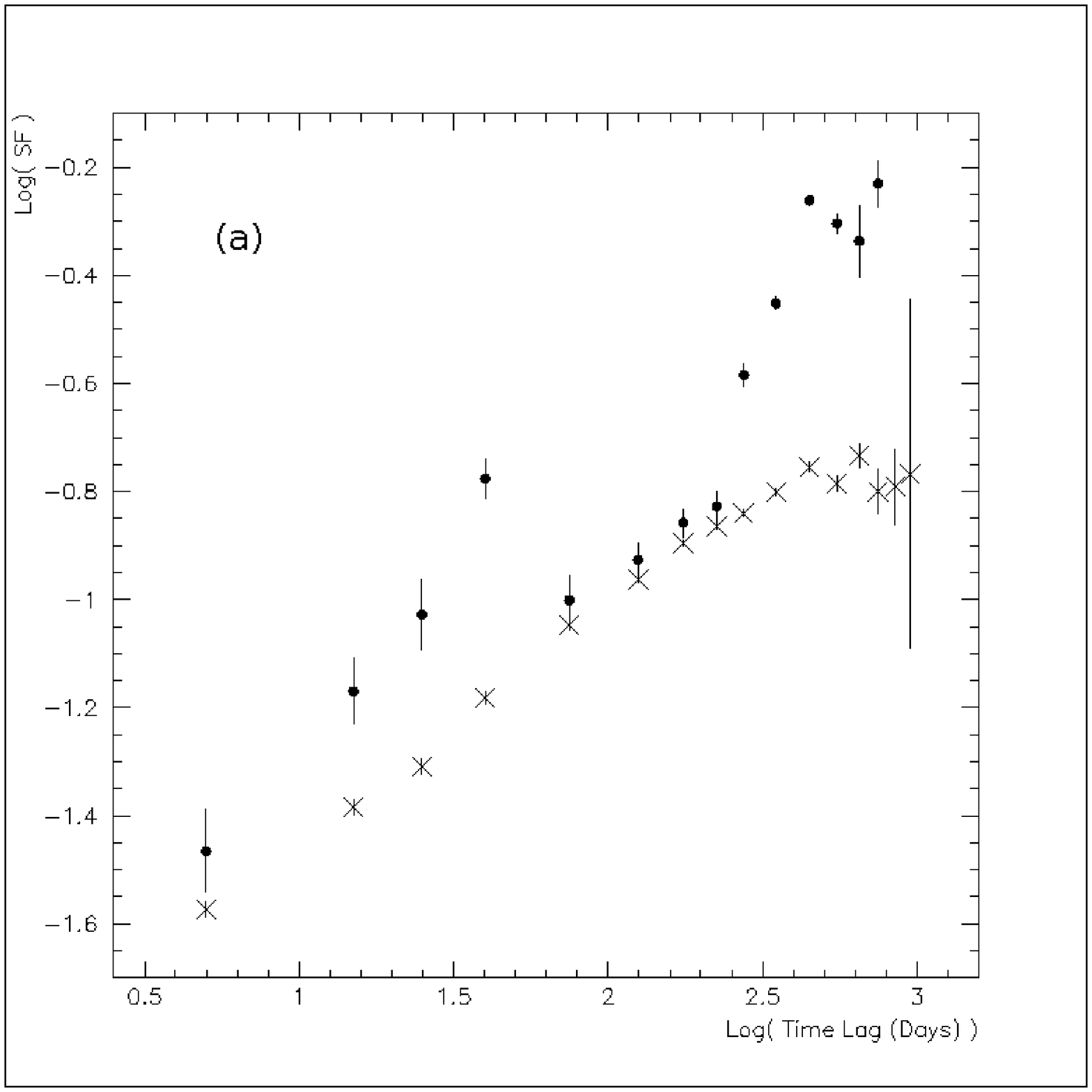}{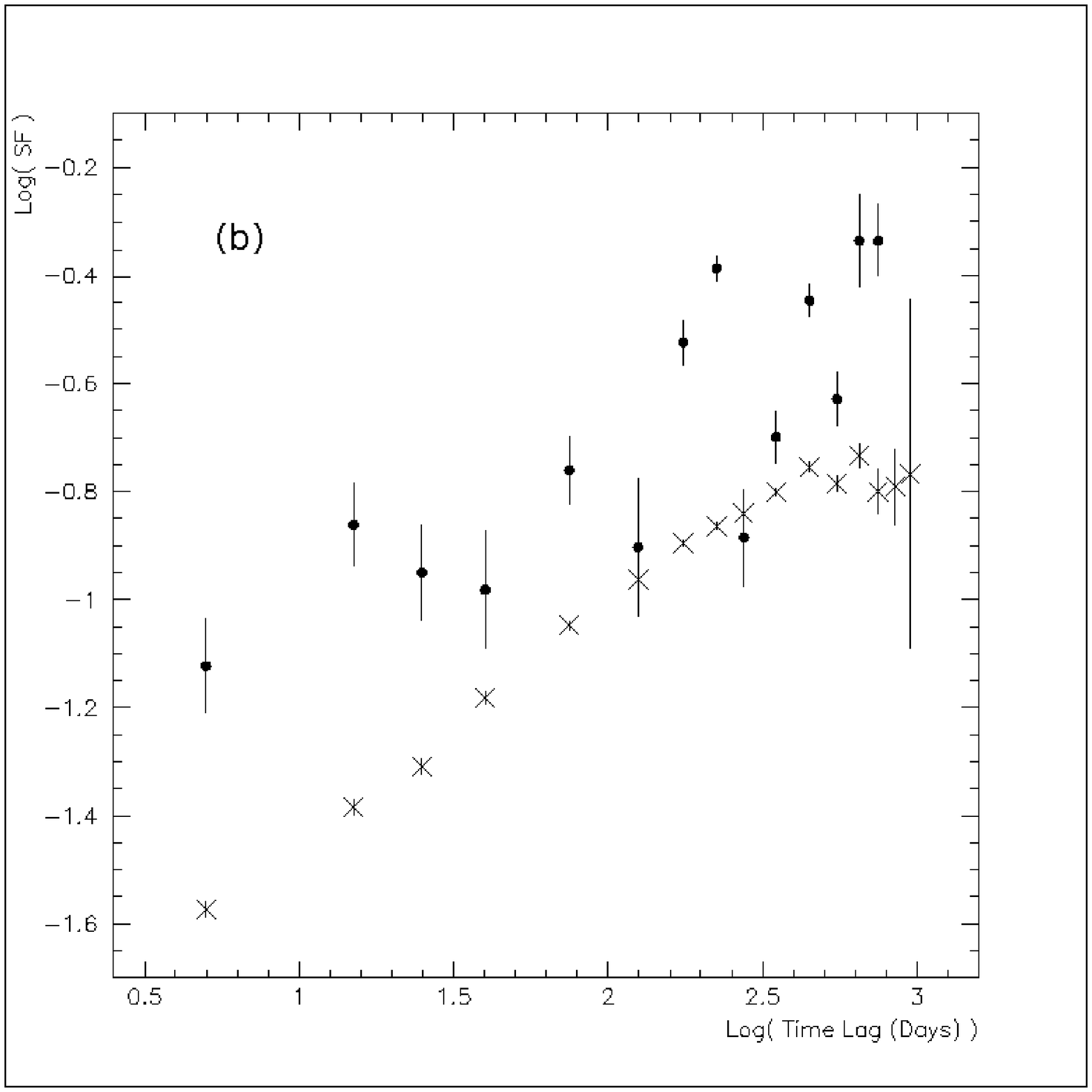}
\caption{Structure function vs. rest frame time lag for FSRQs (a) and BL Lacs (b).  Plotted on a log-log scale.  Filled circles are blazar data, $\times$ symbols are quasar data from Paper I.}
\label{blazar_sfs}
\end{center}
\end{figure}

The structure functions for the FSRQ and BL Lac samples, plotted versus 
rest frame time lag between measurements, are shown as the 
solid points in panels (a) and (b) of figure \ref{blazar_sfs}.  
The errors on the structure function points are the $1\sigma$ 
standard deviation of values obtained by analyzing subsets of the 
total sample.  The structure function
 shapes are qualitatively different from that of the type I quasar structure 
function, measured in Paper I using Palomar-QUEST and shown 
for reference as the $\times$s in figure \ref{blazar_sfs}.  
The turnover of the quasar structure function at the longest time lags is 
a known windowing effect, discussed in detail in appendix A of 
Paper I.  The steady rise of the quasar structure function up 
to the windowing turnover indicates a lack of 
characteristic variability timescale in the quasar data.  This is not the 
case for either the FSRQs or the BL Lacs, which show distinct peaks and 
changes in shape.  The FSRQ and quasar structure functions are consistent, 
however, between timescales of roughly 70 and 250 days.  
The structure functions further display the results seen in the $V$ 
histograms, that both BL Lacs and 
FSRQs are at least as variable as type I quasars on all timescales 
measured by the survey.

Because the structure function presents the mean variability of the 
ensemble, it is sensitive to outlying data points.  The data undergo 
three iterations of 3$\sigma$ clipping prior to the averaging, but there 
is no requirement that the result converge.  Because blazars have 
significant high-variability tails, the structure function is not the 
ideal quantity to describe the variability, as it can easily be skewed by 
statistical variations of a poorly sampled, high $\Delta m$ tail.  
Nevertheless, as the structure function has been used to study type 
I quasars as well as individual blazars (e.g. \cite{heidt96}; 
\cite{kartaltepe07}), 
we present the results here to allow for comparisons with other work.

\subsection{SF($\tau$) Modelling}

\begin{figure}
\begin{center}
\plottwo{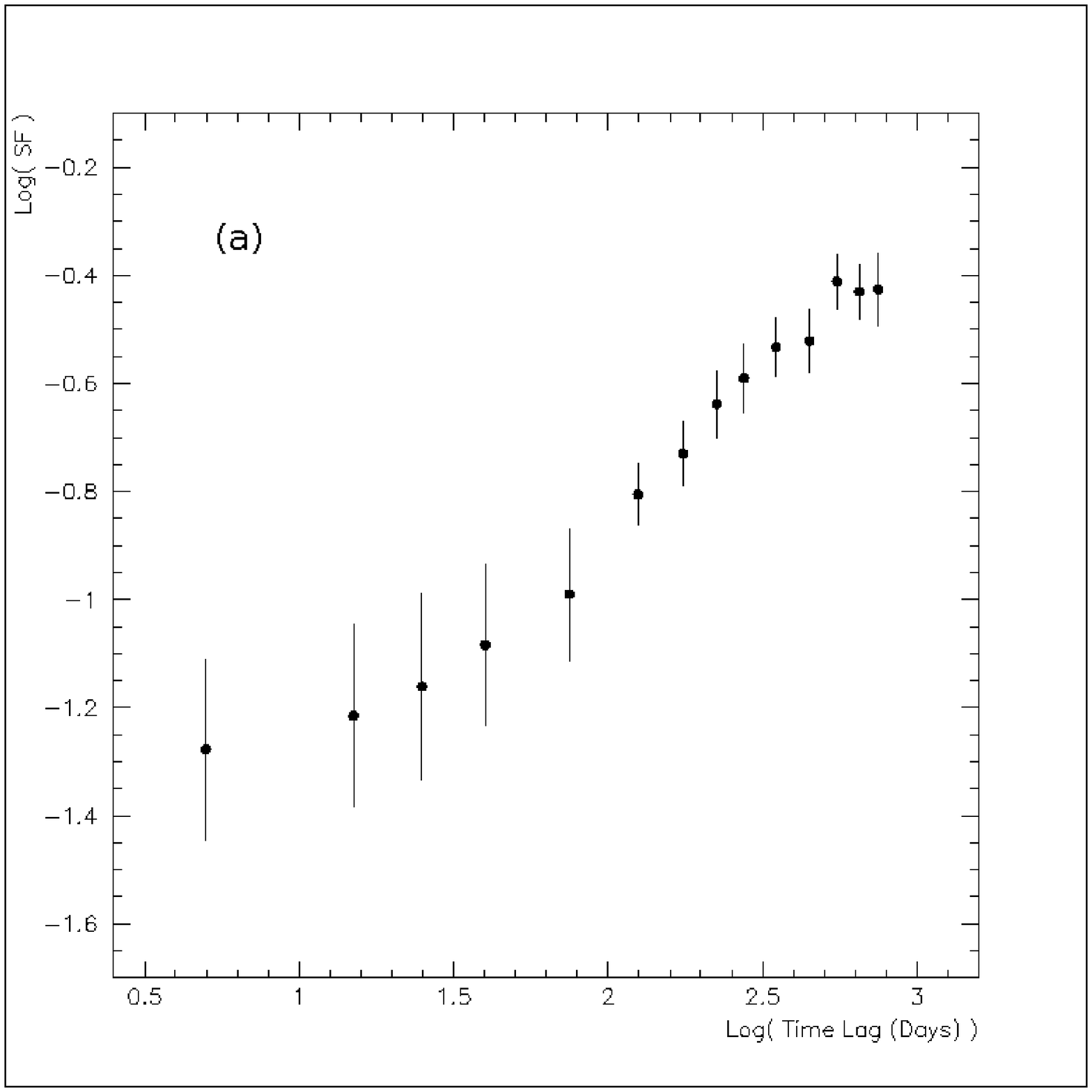}{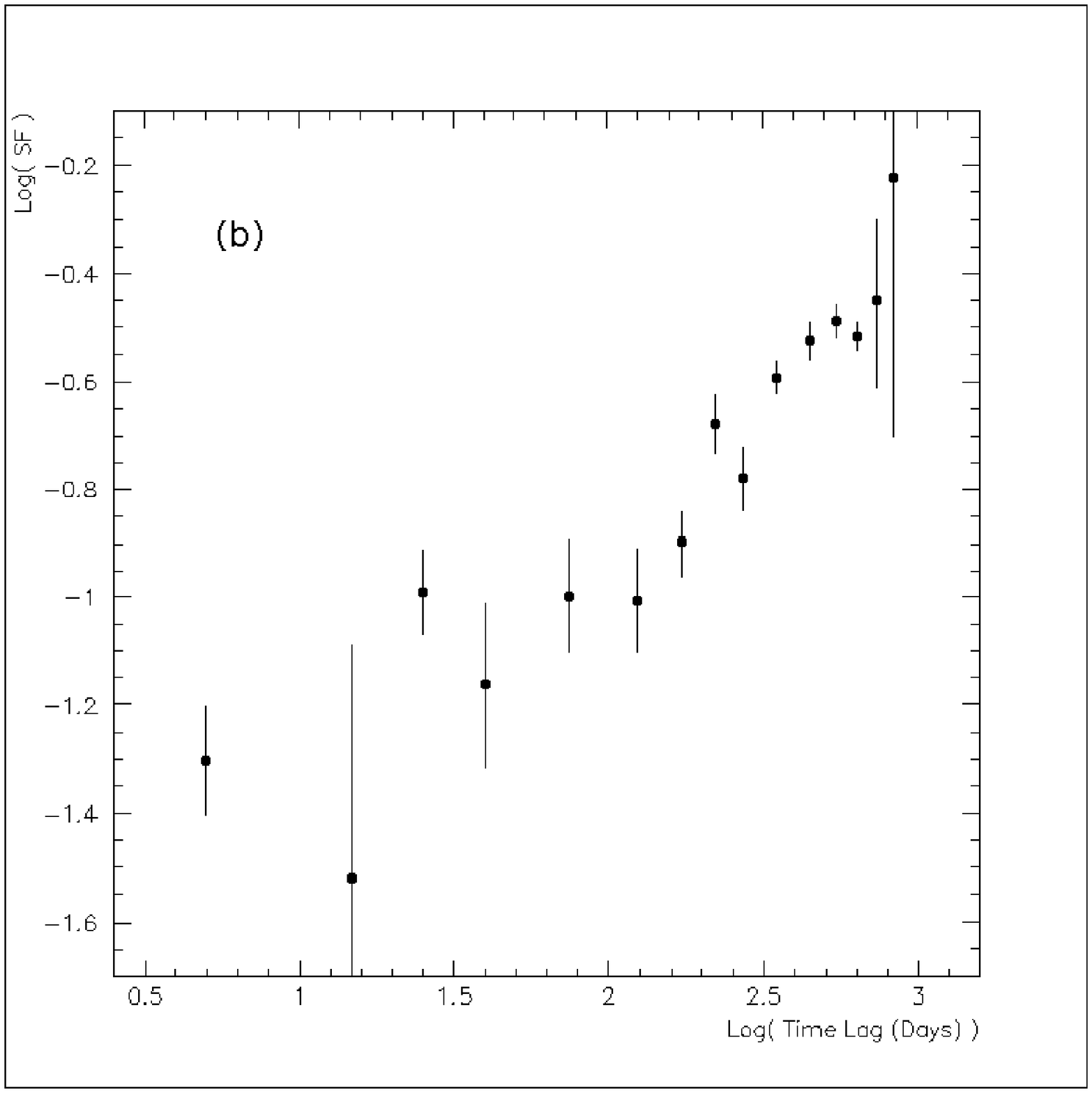}
\caption{Simulated FSRQ (a) and BL Lac (b) structure functions using $PSD \propto f^{-1.71}$ frequency dependence}
\label{sf_171}
\end{center}
\end{figure}

To demonstrate that the difference seen between the structure 
functions of quasars 
and blazars is not due to the data statistics and cadence, we generate 
lightcurves with power spectral distribution 
(PSD) proportional to $f^{-1.71}$, which fit the quasar results well in 
Paper I.  We then analyze the model lightcurves using only simulated 
data points from epochs for which we have real data.  
The resulting simulated structure functions, thereby 
subjected to our blazar window functions, are shown in figure 
\ref{sf_171}.  As expected, the simulated structure functions rise roughly 
as a power law, with plateaus at low and high time lags due to 
measurement noise and edge effects.  
The low statistics cause the BL Lac simulated result to be noisy, 
yet it is consistent with the expected shape.  

These simulated structure functions are qualitatively unlike the data; 
a power law frequency dependence does 
not yield the peaks and changes in slope observed in the blazar 
results.  
In fact, assigning one model lightcurve to every blazar consistently 
produces a structure function with gradual rises and without discrete peaks.  
For example, a lightcurve model 
with $\sim$40 day flares yields a structure function that rises until 
$\tau \sim40$ and remains constant thereafter.  If additional, longer term 
flares are added, the structure function continues rising, with a change of 
slope around 40 days but no discrete peak.  
The peaky, choppy nature of the 
FSRQ and BL Lac structure functions indicates that the variability 
characteristics differ significantly across the sample.  This can be 
due to individual objects behaving uniquely, therefore making up an 
ensemble with a range of typical behaviors.  Or, the mean variability 
can be disproportionately affected by a few measurements with 
high variability amplitude.  
Simulations 
using lightcurve features such as fast shifts in amplitude for only some 
objects do indeed create noisy, peaked structure functions;  however the 
simulation results vary greatly across instantiations.  
Because each blazar is observed typically only four times, 
it is not possible to measure specific lightcurve features directly 
from the data.  
Furthermore, the complicated structure functions do not allow us to 
identify blazar lightcurve 
features such as the flares' exact durations or recurrence rates.  
Instead, the modelling shows us that the blazar lightcurves 
show average variability that is indeed larger than that of type I quasars 
on both short and long timescales, and that the blazars' fluctuations 
over the timescales measured cover a wide range of amplitudes.

\begin{figure}
\begin{center}
\plottwo{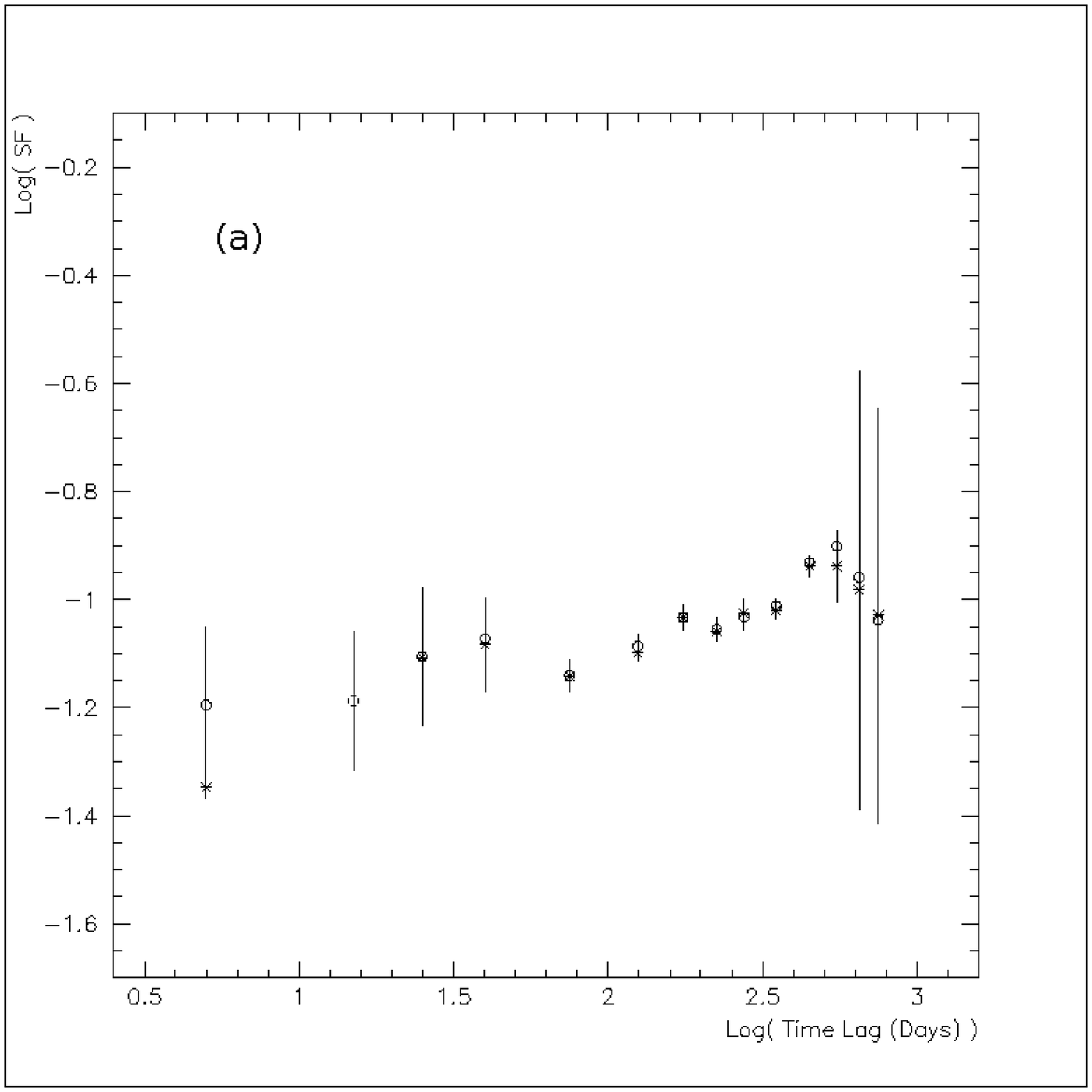}{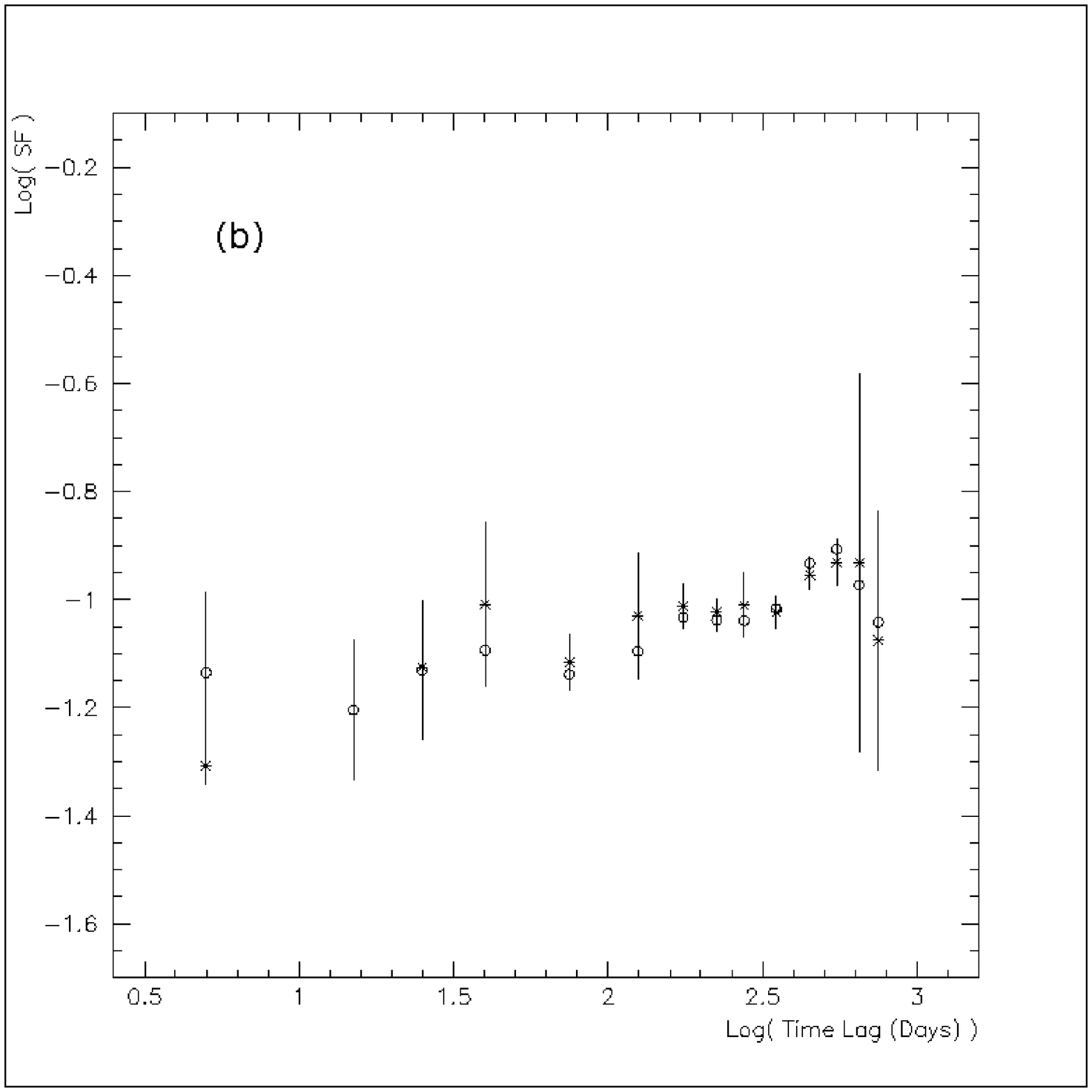}
\caption{Structure functions for 3C 273 using evenly sampled (hollow circles) and QUEST sampled (*) data. (a): FSRQ sampling; (b): BL Lac sampling.}
\label{3c273sf_sampled}
\end{center}
\end{figure}

We can further investigate the influence of the window function  
through the structure function of 3C 273.  We 
use the same method as in section \ref{v_sampling_section} to 
generate evenly-sampled and QUEST-sampled datasets for 3C 273.  
The structure functions calculated from these datasets are shown 
in figure \ref{3c273sf_sampled}, using the FSRQ sampling cadence in 
panel (a) and the BL Lac cadence in panel (b).  The hollow circles 
indicate the evenly-sampled data;  the asterisks indicate the 
QUEST-sampled data.  Error bars are 
determined using subsets of the total sample, as in the main analysis.  
The structure functions are consistent with each other, showing that 
windowing does not significantly alter the structure function results 
for the blazar.

Incidentally, 3C 273 is an example of an individual blazar that 
does not adhere to the description of its class's ensemble variability.  
This object typically does show strong short timescale variability, 
but over the timescales studied its average variability amplitude 
does not increase dramatically with $\tau$.

\subsection{Effects of Radio Jet Velocities}

\begin{figure}
\begin{center}
\plotone{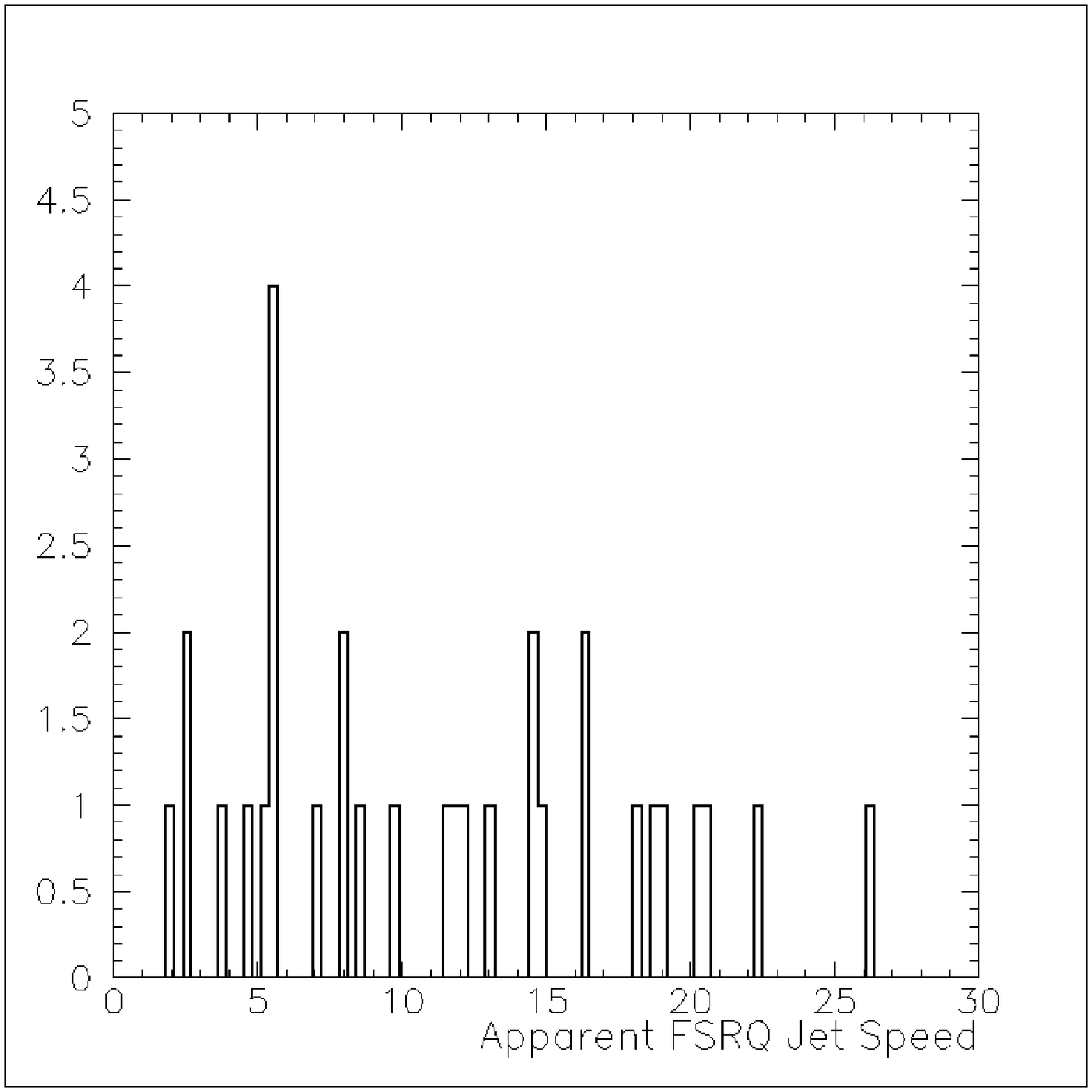}
\caption{$\beta_{app}$ for 31 FSRQs, measured by MOJAVE.}
\label{betahist}
\end{center}
\end{figure}

MOJAVE (\cite{mojave}) has monitored hundreds of blazars' radio 
brightnesses and polarizations using the VLBA.  Their sample includes 
5 of our BL Lacs and 31 of our FSRQs.  
The maximum apparent FSRQ jet speeds $\beta_{app}$ measured by 
MOJAVE\footnote{Data available at http://www.physics.purdue.edu/astro/MOJAVE/} 
are shown in figure \ref{betahist} in units 
of the speed of light.  
Table \ref{vfor3bs_table} lists the median $V$ values for FSRQs with 
different ranges of $\beta_{app}$.  
The range of $V$ increases with $\beta_{app}$, with FSRQs with higher 
$\beta_{app}$ showing more high amplitude variability.

\begin{table}
\begin{center}
\begin{tabular}{|l|l|}
\hline
$\beta_{app}$ & Median $V$ \\ \hline
0 to 7 & 0.09 $\pm$ 0.06 \\
7 to 18 & 0.17 $\pm$ 0.12 \\
18 to 27 & 0.30 $\pm$ 0.25 \\ \hline
\end{tabular}
\caption{Median variability $V$ for FSRQs with $0 < \beta_{app} \leq 7, 7 < \beta_{app} \leq 18$, and $18 < \beta_{app} \leq 27$.}
\label{vfor3bs_table}
\end{center}
\end{table}

\subsection{Variability of CRATES Objects}

\begin{figure}
\begin{center}
\plotone{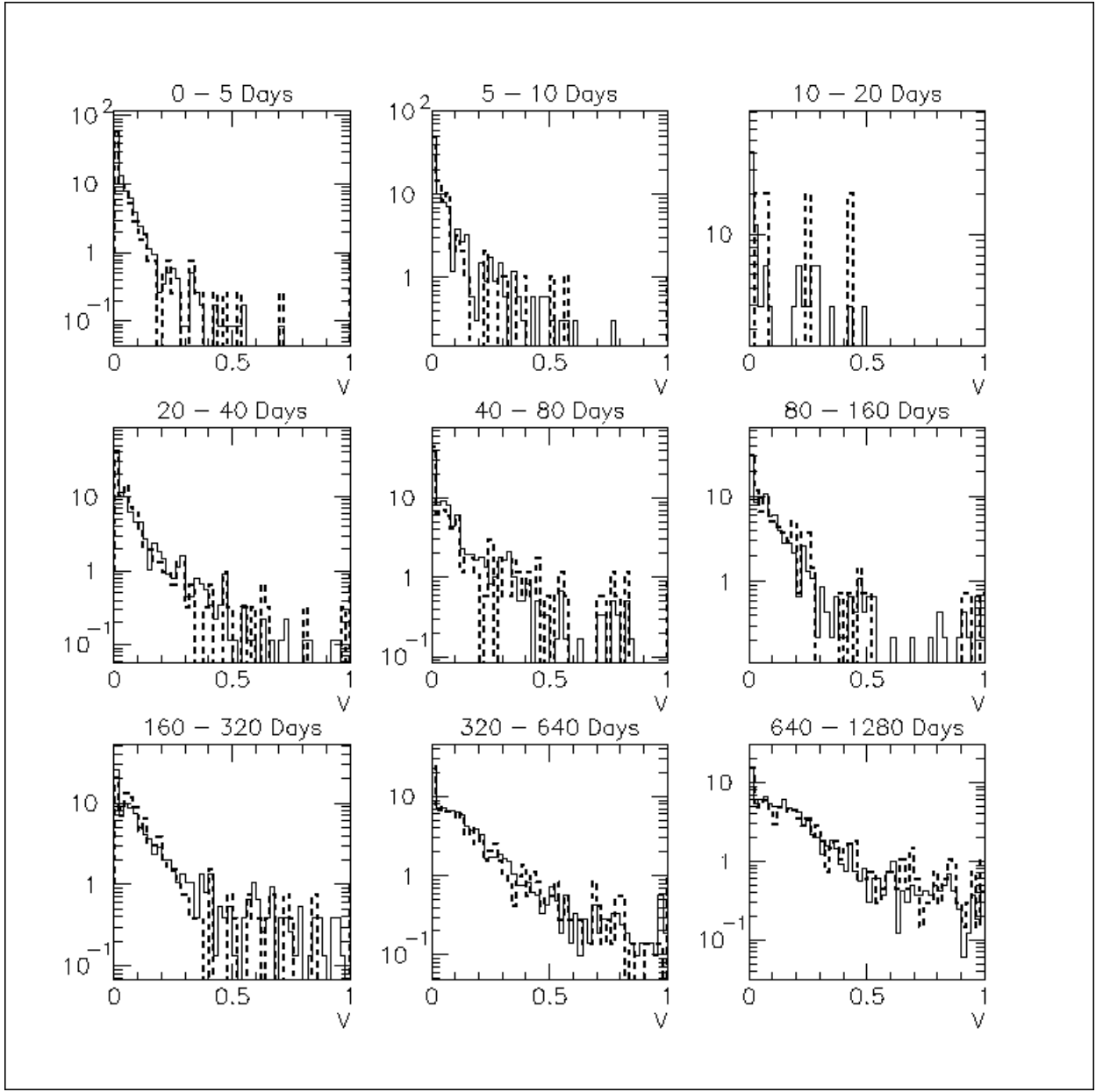}
\caption{Variability $V$ histograms of CRATES objects (solid line) and FSRQs evaluated at z=0 (dashed line) for 9 different time lag bins.}
\label{crates_figure}
\end{center}
\end{figure}

CRATES (\cite{crates}) presents a very large sample ($\sim$11,000) 
of FSRQ 
candidates selected in the 8.4 GHz radio band.  
We compare the optical variability of these candidates with that of 
the confirmed FSRQ sample.  The samples differ in two important 
ways:  the CRATES candidates have unknown redshifts and include fainter 
objects than the FSRQs.  To compare variability timescales, 
we use the observer 
frame time lag $t$ rather than $\tau$ and recalculate the variability 
distributions for the known FSRQs also using the observer's frame time 
lag between measurements.  
If the distributions of object redshifts are similar between the 
two groups then the variability behavior of the ensembles in the 
observer's frame will be directly comparable.  The difference in 
magnitude range between the samples affects $V$ because the 
measurement error $\sigma$ is dependent on the object magnitude.  We 
therefore do not compare all of the CRATES candidates covered by 
Palomar-QUEST (1,917 objects) with all of the 278 FSRQs, but 
subsets of each with statistically equivalent magnitude distributions:  
866 CRATES candidates and 260 FSRQs.  The 
$V$ distributions of these subsets are shown in figure \ref{crates_figure}, 
with the CRATES objects shown as solid lines and the FSRQs as dashed 
lines.  The time lag bin size increases with time lag so that the 
longer time lag histograms are minimally sensitive to small 
differences in the objects' redshift distributions.  
The plots are each normalized to a total of 100 objects to 
facilitate comparisons.  
Clearly, the variability timescales and amplitudes of the CRATES objects 
and the FSRQ sample are qualitatively similar.  To quantify whether the 
$V$s of the two samples come from the same underlying distribution, 
we compare the data using a K-S test for each time lag interval.  
Because the K-S test is sensitive to the shape of the distribution, we 
ignore datapoints with $[m(t) - m(t-\tau)]^{2} < \sigma^{2}$ since our 
arbitrary assignment of $V=0$ for these measurements artificially 
modifies the distribution.  The probabilities that the $V$ data 
come from the same distributions range from 4\% to 26\%, with a
median of 12\%.  The optical variability behavior 
of the CRATES objects and the FSRQs is therefore similar on all 
timescales measured.  Any differences in the redshift distributions 
between the CRATES and the FSRQ samples will cause the 
distributions to differ, so the measured probabilities can 
be taken as lower limits.
The similarity we see in optical variability properties between the CRATES 
sample and known FSRQs provides evidence that the CRATES objects are dominated 
by FSRQs similar to those found using a variety of radio and X-ray methods.

The CRATES sample has a small but significant overlap with objects 
spectroscopically identified as quasars by SDSS.  The quasar sample 
used in figures \ref{fsrqqso}, \ref{bllacqso}, and \ref{blazar_sfs}, 
and described in detail in Paper I, is comprised of $\sim$23,000 
predominantly UV-selected objects from SDSS Data Release 5 
(\cite{sdss5}).  The objects are optically spectroscopically 
confirmed to be quasars;  however the presence of broad emission lines in 
both type I quasars and FSRQs can lead to overlapping identifications.  
There are 83 objects common to the CRATES and type I quasar samples.  
Thirty-eight additional CRATES objects are in the SDSS DR5 spectroscopic quasar 
list, but were removed from the Paper I data set due to either 
their presence in other blazar samples (36 objects) or their low redshift 
(2 objects).  Therefore, 6\% of the CRATES sample observed by QUEST 
has been seen by SDSS to have quasar-like optical colors and spectra.  
This does not disqualify the objects from being FSRQs, but 
illustrates the difficulty of distinguishing between AGN classes and 
highlights the influence of selection effects on samples of AGN.  
The CRATES objects make up 0.4\% of the quasar 
sample used in this work and Paper 1.  This order of 
contamination will not affect the average variability properties of 
the type I quasar ensemble.

\subsection{Variability Duty Cycle}

\begin{figure}
\begin{center}
\plotone{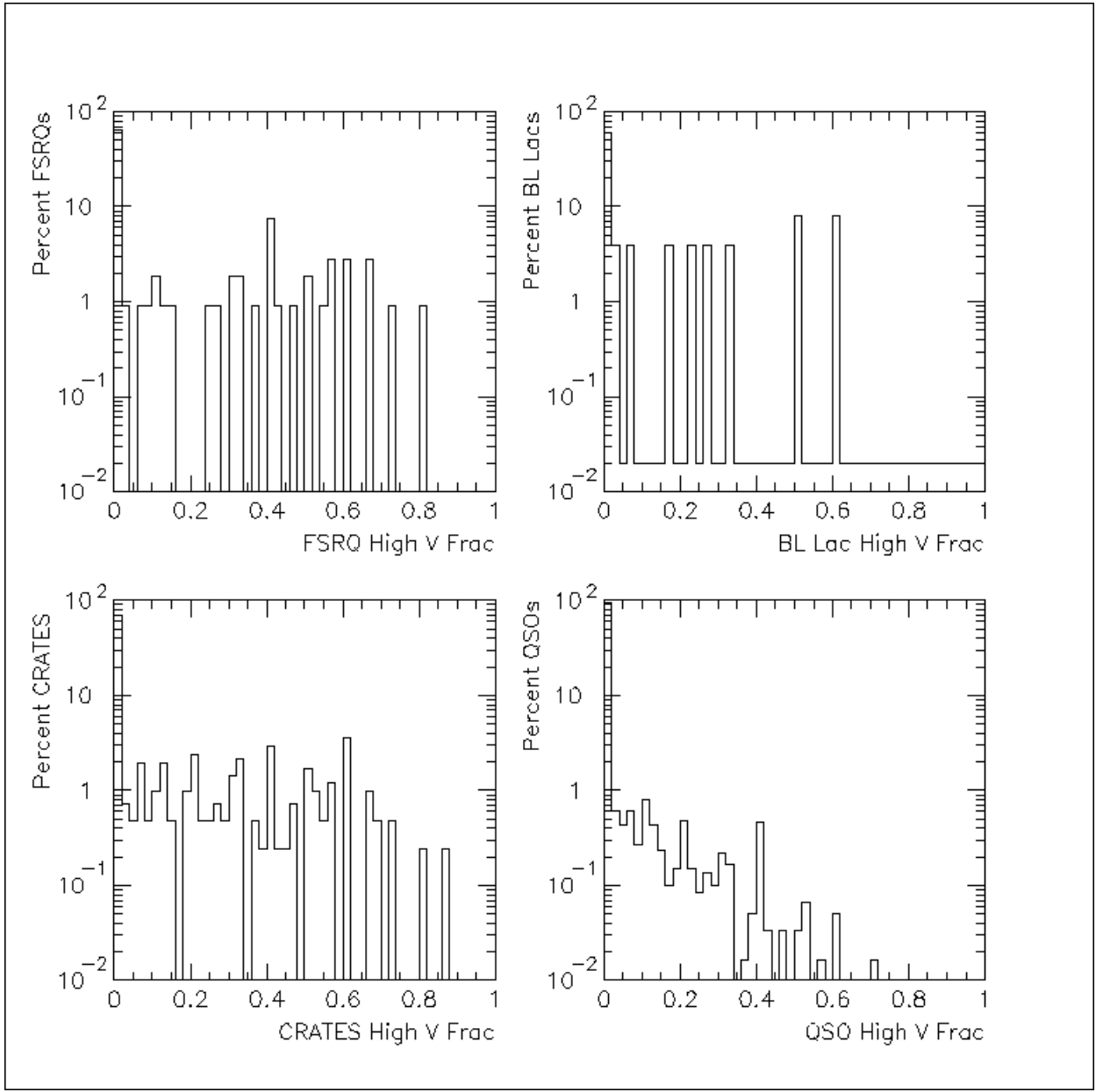}
\caption{Fraction of $V$ measurements greater than 0.4, for objects with at least 10 measurements.  Number of objects on the Y axis is normalized to a total of 100 in each plot.}
\label{fracgt04}
\end{center}
\end{figure}

About 15\% of the blazars' $V$ measurements are greater than 0.4.  
Are these measurements due to each object varying dramatically during 
15\% of its measurements, or a small fraction of the blazars varying 
with high amplitude all the time, or something 
between?  Figure \ref{fracgt04} shows, for FSRQ, BL Lac, CRATES, 
and quasar samples including only objects with at least 10 $V$ 
measurements, the fraction of an object's $V$ measurements that are 
greater than 0.4.  Each histogram is normalized to a total of 100 
objects.  
60\%, 65\%, and 69\% of the BL Lacs, FSRQs, and CRATES objects, 
respectively, have no $V$ values greater than 0.4.  Those for which 
we do measure such 
high variability are evenly distributed up to a measurement fraction 
of roughly 0.8:  e.g. equal numbers of blazars are highly variable 
across 10\% of our measurements 
as are highly variable across 70\% of them.  On the other 
hand, we see 94\% of quasars always with $V \leq 0.4$;  those 
which do appear highly variable tend to show such variability 
across fewer than $\sim40\%$ of their measurements.  
These results reinforce the conclusion that blazars have a large range 
of behavior;  they show active and quiescent phases on all timsecales 
spanned by our survey.  The majority of blazars, however do not 
contribute to the highest levels of variability seen by the survey.

We caution that without close, even spacing of observations it is 
not possible to determine objects' intrinsic duty cycles.  By 
selecting objects with at least 10 measurements we ensure that the 
blazars studied here are observed on a range of timescales over at 
least two years.  However, by the nature of a ground based survey, 
there are months at a time during which we have no observations of a 
particular blazar.  We therefore cannot say that a blazar varies 90\% 
of the time, only over 90\% of our measurements.  This analysis 
describes the variety of variability behavior observed in blazars 
over timescales from hours to several years, and is the best approximation 
to an intrinsic duty cycle measurement as can be made with such a 
survey.

\section{Discussion \label{discussion_section}}

Previous variability studies of blazars have used many, frequent 
measurements of one or a few objects to examine the details 
of individual fluctuations in the sources.  In this work we 
consider a few measurements of many blazars in order to study 
the common variability properties of the class as a whole.  
Our results reflect the ensemble behavior of a diverse population.  
The choppy nature of the FSRQ and BL Lac structure functions 
suggests that blazar variability differs widely from object to object.  
This is also shown by the wide range of high variability duty cycles 
seen in the objects.  About 35\% of the blazars are seen to vary by 
$V > 0.4$ during the 3.5 years of the Palomar-QUEST survey.  
These highly variable objects have a roughly continuous 
range of high variability duty cycles up to about 80\% of the 
measurements, showing that while some blazars vary drastically 
over nearly all observations, others show 
active periods with no single characteristic length within the time 
span of our survey.  Because our blazar samples show such diversity, 
the ensemble variability properties cannot be interpreted as 
describing the behavior of individual objects.  Instead, they highlight 
variability amplitudes and timescales that are common in BL Lacs and 
FSRQs, and illustrate differences in typical behaviors between classes.

We see that blazars show some variability properties similar 
to those of type I quasars.  Specifically, the majority of the 
variability takes place at amplitudes below a few tenths of a 
magnitude, and the low-amplitude peak of variability broadens 
with increasing time lag.  Both FSRQs and BL Lacs show a 
higher amplitude tail superimposed on this low amplitude 
peak;  in FSRQs the quasar-like peak is more prominent.  
This similarity between FSRQs and quasars may be related to their 
spectral similarity:  the presence of thermal emission, including broad 
emission lines.  Accretion disk flux, partly reprocessed in the 
broad line region, may be responsible for much of the low amplitude 
variability seen in FSRQs.  
The flatter shape of the BL Lac variability distribution compared 
to the FSRQs suggests 
that while accretion disk emission may also play an important 
role in BL Lac variability, its contribution is less significant than 
in FSRQs.  

The highest amplitude variability, with $V \gtrsim$ 1 
magnitude, is more common in FSRQs than BL Lacs at all timescales.  
High amplitude blazar variability is commonly attributed to shock 
dissipation in the jet;  the fact that the strongest variability 
differs between FSRQs and BL Lacs implies different jet 
characteristics for the two groups.  Indeed, there is strong 
evidence, independent of variability, that FSRQs and BL Lacs have 
different jet energies and morphologies: BL Lacs are thought to be 
intrinsically FR I radio galaxies, and FSRQs to be FR IIs.  
The higher energy jets characteristic of FR IIs may 
be important in generating the excess high amplitude variability 
observed in FSRQs.

Blazars exhibit a wide variety of jet energies, with differences due to 
intrinsic properties as well as beaming direction.  For example, the apparent 
velocity of blazar jets in the radio band has been seen to range from 
stationary to over $\sim 30$ times the speed of light! (\cite{cohen07})  
31 of the FSRQs in our sample have been monitored by MOJAVE and 
have maximum jet speed measurements.  Those FSRQs with faster jets 
show more high-amplitude variability than those with slower jets.  
This could be because the faster radio regions 
generate shocks which create more optical flux;  in this case the radio and 
optical properties would be directly related, generated in the same 
region of the jet.  Or, the radio and optical flux could be indirectly 
related, with the powerful injections which fuel the radio speed also 
enhancing optical variability elsewhere in the jet.

The structure function is essentially the clipped 
mean of the variability amplitude histogram, plotted versus rest 
frame time lag, and has often been used to study AGN variability.  
The FSRQ and BL Lac structure functions have 
shapes unlike that of type I quasars, showing 
excess variability on both short and long timescales.  
The blazars' structure functions reinforce the 
conclusion that FSRQ variability is more similar to quasars' than 
BL Lac variability is to quasars', particularly at timescales 
of roughly 70 to 250 days.  
The FSRQ structure function shows a clear excess of power compared to type 
I quasars around timescales of 20 to 50 days.  This detection is consistent 
with reports of flares with similar timescale in objects like GC 0109+224 
(\cite{ciprini03}).  However, this power 
excess can not be a result of flares at that timescale in all blazars, 
as that kind of recurrent, well-behaved fluctuation would consistently 
lead to a smooth, monotonic structure function shape that is qualitatively 
different from the data.  Equivalently, we can see in figure \ref{fsrqqso} 
that on these timescales there is increased high amplitude 
variability.  However, such variability is not absent at other timescales, 
even at those where the FSRQ and quasar structure functions are similar.  

The CRATES Survey (\cite{crates}) has identified a very large number 
of FSRQ candidates by their radio emission.  The variability amplitude 
distributions of these 
candidates as seen in the Palomar-QUEST Survey are similar to those of 
confirmed FSRQs at all measured time lags.  
The CRATES selection criteria are independent of variability; the 
similarity of their optical variability to that of FSRQs therefore 
implies that the CRATES 
objects are likely to be predominantly FSRQs.  These candidates 
promise to be an extremely valuable set from which to study the 
ensemble properties of blazars, particularly if redshifts of the 
objects can be obtained so that more accurate timescales of 
variability may be measured.

For this work we have divided the blazar sample into BL Lacs 
and FSRQs because there are significant differences in the optical spectral 
shape between the two classes, likely related to jet morphology and differences 
in the host radio galaxy.  
It may be helpful to divide the blazar sample 
in different ways that could be more in line with their different variability 
properties.  For example, BL Lacs can be divided into HBL and LBL groups 
(high peaked and low peaked BL Lacs) according to the central frequencies 
and relative heights of the synchrotron and Compton peaks in the 
SED.  FSRQs typically have SED peak frequencies similar to LBLs, 
although some have been seen with SEDs more like HBLs' 
(\cite{padovani07}).  It is not clear how the HBL/LBL 
status of the blazar relates to the object's fundamental properties.  
It has been proposed that HBL jets may be dominated by synchrotron 
self-Compton (SSC) shock dissipation while the LBLs may follow external 
radiation Compton (ERC) mechanisms.  This is supported 
by the fact that FSRQs, which tend to have SEDs similar to LBLs, exhibit 
broad optical emission lines;  the presence of this thermal emission 
is consistent with the ERC model's invocation 
of disk and broad emission line radiation acting as Compton seed photons 
(\cite{fossati98}).  The HBL/LBL dichotomy is one example of a different 
blazar categorization that may yield optimally contrasting, and therefore 
illuminating, 
variability results.  Such rebinning of the data awaits further 
multiwavelength and variability information 
to provide sufficient statistics.

\section{Conclusions \label{conclusions_section}}

We have used the Palomar-QUEST Survey to examine the ensemble 
optical variability of 276 FSRQs and 86 BL Lacs.  This is the first 
study of the common variability characteristics of so many of these 
rare AGN.  We have therefore described the 
variability in several ways, showing distributions of variability 
amplitudes at numerous time lags as well as a more traditional 
structure function analysis.

We find that blazars have a wide range of behavior from object 
to object.  The diversity of behavior within the blazar 
class is qualitatively apparent in the shape of the ensemble structure 
functions.  We also see that the duty cycle of blazars differs 
widely from object to object.  
About the same fraction of objects show high amplitude variability 
across $\sim 5\%$ to $\sim 80\%$ of their measurements.  Roughly 65\% 
of blazars do not show this high amplitude variability 
($\Delta m \gtrsim 0.4$ magnitudes) at all during the survey.  
Therefore, while optical variability may be a distinctive feature of 
blazars, details of the variability appear very different between 
objects on timescales up to a few years.  
The ensemble variability of the class, studied using 
few measurements of many objects, therefore reflects behavior 
characteristic of the group as a whole rather than descriptive of 
individual members.  

Both BL Lac and FSRQ classes have similarities in variability 
amplitudes and timescales compared to type I 
quasars, however with more fluctuations superimposed at all timescales  
measured by Palomar-QUEST.  FSRQs show more similarities to type I quasars 
than BL Lacs do, implying a stronger role for accretion disk variability 
in FSRQs.  The heightened presence of disk variability is in accordance 
with the presence of broad emission lines in FSRQ spectra, which are 
absent in BL Lacs.  Periodicity is not observed to be a common 
variability property in blazars.  

FSRQs exhibit more variability at the highest amplitudes than 
BL Lacs.  This is likely related to higher jet energies in 
FSRQs compared to BL Lacs.  For a subset of 31 FSRQs with measured 
maximum apparent jet speeds in the radio band, the radio jet speed 
correlates positively with the optical variability amplitude.  

CRATES candidate FSRQs show similar variability amplitudes and 
timescales to the confirmed FSRQs.  This behavior is independent 
of the CRATES selection criteria, and is therefore a good test to show 
that this sample is likely dominated by FSRQs similar to those known.

The discovery of new blazars, for example from the CRATES survey, 
is very important for the thorough study of this rare class.  
Characterizing the appearance of blazars in sparsely sampled, wide 
field surveys such as Palomar-QUEST facilitates the optical 
variability-based selection of new blazars.  With more statistics 
we may be able to divide the class into subsets 
that better clarify the processes behind the dramatic variability.

\acknowledgements

We thank S. G. Djorgovski and his group for helpful discussions.  
This work is supported by the Office of Science of the Department of 
Energy and the National Science Foundation.

This research has made use of data from the MOJAVE (Lister and Homan, 
2005, AJ, 130, 1389), 2cm Survey (Kellermann et al., 2004, ApJ, 609, 
539), and CRATES (Healey et al., 2007, ApJS, 171, 61) programs.

{\it Facilities:} \facility{PO:1.2m}

\end{document}